\documentclass[11pt]{article}
\usepackage[a4paper, margin=1in]{geometry}
\usepackage{times}
\usepackage{hyperref}
\usepackage{graphicx}
\usepackage{amsmath}
\usepackage{titlesec}
\usepackage{authblk}
\usepackage{abstract}
\usepackage{cite}
\usepackage{booktabs}
\usepackage{amssymb}
\usepackage{pgfplots} % For progress bars
\usepackage{colortbl} % For table shading
\usepackage{adjustbox} % For resizing the table if necessary
\usepackage[T1]{fontenc}
\usepackage[utf8]{inputenc}
\usepackage{url} % 添加url包支持\url命令

\titleformat{\section}{\large\bfseries}{\thesection}{1em}{}

\title{\textbf{EnTao-GPM: DNA Foundation Model for Predicting the Germline Pathogenic Mutations}}

\author[1,2]{Zekai Lin\thanks{Equal contribution. This research features four co-first authors who contributed equally to this study: Zekai Lin, \url{zklin23@m.fudan.edu.cn}, Haoran Sun,  \url{manglu3935@126.com}, Yucheng Guo, \url{yucheng@biomap.com}, and Yujie Yang, \url{yujie@biomap.com}. Additional contributing authors include Yanwen Wang, \url{24211010074@m.fudan.edu.cn}, Bozhen Hu, \url{bozhen@biomap.com}, Chonghang Ye, \url{chonghang@biomap.com}, and Qirong Yang, \url{qirong@biomap.com}. Correspondence regarding this work should be addressed to the following}}
\author[1,2]{Haoran Sun\textsuperscript{*}}
\author[3]{Yucheng Guo\textsuperscript{*}}
\author[3]{Yujie Yang\textsuperscript{*}}
\author[1,2]{Yanwen Wang}
\author[3]{Bozhen Hu}
\author[3]{Chonghang Ye}
\author[3]{Qirong Yang}
\author[2]{Fan Zhong\thanks{Corresponding author:\url{zonefan@163.com}}}
\author[3]{Xiaoming Zhang\thanks{Corresponding author:\url{zhangxiaoming@biomap.com}} }
\author[2,4,5]{Lei Liu\thanks{Corresponding author:\url{liulei_sibs@163.com}} }
\affil[1]{School of Basic Medical Science, Shanghai Medical College, Fudan University, Shanghai 200032, China}
\affil[2]{Intelligent Medicine Institute, Fudan University, Shanghai 200032, China}
\affil[3]{BioMap Research, Beijing 100086, China}
\affil[4]{Shanghai Institute of Infectious Disease and Biosecurity, Fudan University, Shanghai 200032, China}
\affil[5]{Shanghai Institute of Stem Cell Research and Clinical Translation, Shanghai 200120, China}

\date{July 2025}

\begin{document}
\maketitle

\begin{abstract}
Distinguishing pathogenic mutations from benign polymorphisms remains a critical challenge in precision medicine. EnTao-GPM, developed by Fudan University and BioMap, addresses this through three innovations: (1) Cross-species targeted pre-training on disease-relevant mammalian genomes (human, pig, mouse), leveraging evolutionary conservation to enhance interpretation of pathogenic motifs, particularly in non-coding regions; (2) Germline mutation specialization via fine-tuning on ClinVar and HGMD, improving accuracy for both SNVs and non-SNVs; (3) Interpretable clinical framework integrating DNA sequence embeddings with LLM-based statistical explanations to provide actionable insights. Validated against ClinVar, EnTao-GPM demonstrates superior accuracy in mutation classification. It revolutionizes genetic testing by enabling faster, more accurate, and accessible interpretation for clinical diagnostics (e.g., variant assessment, risk identification, personalized treatment) and research, advancing personalized medicine.
\end{abstract}

% --------------------------------------------------
\section{Introduction}

Interpreting genetic mutations, particularly their association with heritable diseases, remains a critical bottleneck in precision medicine. The exponential growth of genomic sequencing data has outpaced the capacity of traditional bioinformatic tools to accurately distinguish pathogenic mutations from benign polymorphisms\cite{1shendure2012expanding}. In response, large-scale AI models, including DNA foundation models and large language models, have emerged as transformative technologies that enable precise and scalable genomic interpretation.

\subsection{Pathogenicity Prediction via Rule-Based and Data-Driven Models}

Historically, pathogenicity prediction was based on rule-based tools (e.g., SIFT, PolyPhen-2) that scored mutation using evolutionary conservation or amino acid properties\cite{6adzhubei2010method}. While foundational, these tools struggled with noncoding mutations and rare mutations, lacking scalability to large genomic datasets. The shift to data-driven models addressed these limitations, with ClinVar\cite{7Landrum2025clinvar} emerging as a cornerstone resource by curating expert-annotated mutation pathogenicity labels. Models like REVEL\cite{8Ioannidis2016} integrated multiple predictors to improve single nucleotide mutation(SNV) classification, but remained constrained by reliance on hand-made features.

DNA pre-training models revolutionized this space by enabling end-to-end learning of pathogenicity from raw sequences. Evo 2\cite{4brixi2025genome} demonstrated zero-shot prediction of mutational effects across coding and non-coding regions, outperforming REVEL in cross-species generalization. Similarly, DNABERT-2\cite{9zhou2023dnabert}, a transformer model fine-tuned on ClinVar, showed improved accuracy in distinguishing pathogenic single nucleotide mutations(SNVs) by leveraging contextual sequence embeddings. Notably, TrinityDNA\cite{10yang2025trinitydnabioinspiredfoundationalmodel} advanced non-SNV prediction by modeling insertions/deletions with variable-length sequence inputs, achieving state-of-the-art performance on indel pathogenicity tasks.

To enrich pathogenicity training data, Human Gene Mutation Database (HGMD)\cite{11stenson2020human} complemented ClinVar with high-confidence disease-causing mutations, enhancing the generalization of the model across mutation types, critical for comprehensive clinical sequencing reports.

\subsection{DNA Pre-training Models}
A foundational breakthrough in genomic AI has been the development of DNA pre-training models, which learn universal sequence features from large-scale genomic data to enable downstream task generalization. Early efforts, such as DNABERT\cite{2ji2021dnabert}, demonstrated that transformer architectures could capture nucleotide-level contextual patterns, outperforming traditional k-mer methods in tasks like promoter prediction. However, these models were limited by narrow training datasets (e.g., human-centric) and shallow sequence depth, restricting their ability to generalize across species or complex mutation types.

Subsequent models expanded both the data scale and the biological scope. Evo\cite{3nguyen2024sequence}, trained on 2.7 million prokaryotic genomes, pioneered the prediction of single nucleotide resolution of mutation impacts on molecular and genomic scales, establishing that deep learning could link sequence variation to organismal fitness. Its successor, Evo 2\cite{4brixi2025genome}, further advanced this paradigm by leveraging 9.3 trillion nucleotides spanning all domains of life, integrating a StripedHyena architecture to enable zero-shot prediction of mutational effects in noncoding regions, a critical capability for interpreting the 98\% of the human genome not encoding proteins. TrinityDNA\cite{10yang2025trinitydnabioinspiredfoundationalmodel}, a bio-inspired foundational model, pushed this frontier further by integrating biological structural insights and multi-scale sequence modeling. It introduced Groove Fusion modules to capture DNA's major and minor groove features, Gated Reverse Complement mechanisms to leverage strand symmetry, and a sliding multi-window attention to balance local and global dependencies—addressing the locality bias of structured state space models and oversmoothing in full attention. Its evolutionary training strategy, progressing from prokaryotic (average 924 bp) to eukaryotic genomes (tens of kilobases) with expanding context windows (8k to 100k), enhanced cross-species generalization, outperforming EVO and Caduceus in zero-shot tasks across prokaryotic and eukaryotic organisms, and setting new benchmarks in CDS annotation with superior precision and F1 scores.

Parallel progress came from LucaOne\cite{5He2024.05.10.592927}, a model trained on 169,861 species to unify nucleic acid and protein sequence learning. By integrating DNA, RNA, and protein modalities, LucaOne demonstrated that cross-modal pre-training improves tasks like DNA-protein translation and regulatory element prediction. This work validated the value of multi-omics integration for genomic interpretation, laying groundwork for models that bridge sequence conservation and functional relevance.

\subsection{Large Language Models in Biomedical Interpretation}

LLMs have increasingly been integrated into genomic interpretation, bridging sequence data and clinical meaning. GPT-4\cite{12achiam2023gpt} demonstrated utility in synthesizing the biomedical literature to explain the associations of mutation diseases, while BioBERT\cite{13lee2020biobert} specialized in the extraction of biomedical text, allowing the extraction of mutation pathogenicity relationships from unstructured data. Med-PaLM 2\cite{14singhal2025toward} further advanced this by showing high accuracy in interpreting clinical reports, setting a precedent for AI-driven clinical decision support. However, few models have been tailored to provide statistical interpretations of genetic sequencing reports—an unmet need addressed by specialized genomic-LLM integrations.

\subsection{Multi-Species Genomics and Generalization for Robust Prediction}

A critical challenge in pathogenicity prediction is generalizing to rare or understudied Mutation, which often lack human-specific data. Multi-species genomic models address this by leveraging evolutionary conservation—Mutation in conserved regions are more likely to be functionally impactful. LucaOne led this effort by training on 169k species, showing that multi-species embeddings improve prediction of human mutation effects by capturing deep evolutionary patterns\cite{5He2024.05.10.592927}. Similarly, Evo 2 used 9.3 trillion nucleotides from all life domains to model conservation, enabling accurate zero-shot predictions in non-model organisms\cite{4brixi2025genome}. DNABERT-2 further advances this paradigm by pre-training on 135 species spanning 6 taxonomic categories (including mammals, fungi, and bacteria) with 32.49B nucleotides, demonstrating that multi-species training enhances performance across diverse genomic tasks while maintaining efficiency\cite{9zhou2023dnabert}. Its architecture, optimized with Byte Pair Encoding (BPE) tokenization and Attention with Linear Biases (ALiBi), achieves comparable performance to state-of-the-art models with 21× fewer parameters, making it suitable for downstream tasks requiring cross-species generalization. Focusing on mammals (human, pig, mouse) commonly used in disease modeling has been shown to balance breadth and relevance, as their genomic conservation more directly maps to human pathogenicity.

\subsection{The EnTao Germline Pathogenic Mutation Foundation Model}

Building on the advancements in DNA foundation models, pathogenicity prediction, and LLM-driven clinical interpretation outlined above, the \textbf{EnTao-GPM (Enlightened Tao Germline Pathogenic Mutations) } model represents a significant advancement in using artificial intelligence to predict the pathogenicity of genetic mutations. As part of the growing field of medical AI, EnTao-GPM focuses on improving the accuracy of genetic mutation interpretation by leveraging deep learning techniques, addressing critical gaps in existing methodologies.

The core innovation of EnTao-GPM lies in its ability to interpret genetic data at a deeper level than previous models, thanks to its training in high-quality genomic sequences and its use of a pre-trained DNA model. The model processes not only SNVs but also non-SNVs, such as insertions and deletions, providing a comprehensive tool for genetic diagnostics. The performance of the model has been validated through extensive testing in authoritative databases such as ClinVar\cite{7Landrum2025clinvar}, and it demonstrates superior accuracy in distinguishing between pathogenic and benign mutations. This enhanced prediction capability has the potential to revolutionize genetic testing, making it more efficient, precise, and accessible for both clinical applications and research.

EnTao-GPM is poised to support clinical professionals in the interpretation of genetic sequencing, where it can help identify risk factors for genetic diseases, assist in patient-specific disease assessments, and guide treatment decisions. By improving the precision and speed of genetic diagnostics, EnTao-GPM aims to be a critical tool in personalized medicine, making it possible to better predict disease risks and optimize treatment strategies based on genetic information.

EnTao-GPM advances existing frameworks through \textbf{three key innovations}:
\begin{enumerate}
    \item \textbf{Targeted Pre-training on Mammalian Genomes} \\
    By focusing on human and the other 26 laboratory mammals relevant to human diseases, EnTao-GPM builds on the cross-species conservation principle validated by LucaOne\cite{5He2024.05.10.592927} and Evo 2\cite{4brixi2025genome}, but with a sharper focus on mammalian genomes. This targeted approach enhances the model's ability to interpret conserved pathogenic motifs, especially in non-coding regions.
    
    \item \textbf{Specialization in Germline Mutation} \\
    EnTao-GPM focuses on germline mutations, which are critical for assessing hereditary disease risks but have been understudied in general genomic models. By fine-tuning on curated datasets such as ClinVar \cite{7Landrum2025clinvar}and HGMD\cite{11stenson2020human}, EnTao-GPM improves its accuracy in distinguishing pathogenic from benign mutations across both SNVs and non-SNVs, addressing limitations that focus on broader genomic data.
    
    \item \textbf{Integration of DNA Sequence Embeddings with LLM-based Statistical Explanations} \\
    Unlike traditional models, EnTao-GPM integrates DNA sequence embeddings with LLM modules to provide actionable clinical insights. The model's dual capability enables it to output detailed statistical interpretations of genetic mutations, providing precise and actionable reports to clinical professionals.
\end{enumerate}

The performance of EnTao-GPM has been validated through extensive testing on authoritative databases such as ClinVar, where it has demonstrated superior accuracy in distinguishing pathogenic from benign mutations. This advanced prediction capability promises to revolutionize genetic testing, making it faster, more accurate, and more accessible for both clinical use and research applications.\cite{19richards2015standards}

EnTao-GPM will support clinicians in interpreting genetic sequencing, aiding in identifying disease risk factors, conducting patient-specific assessments, and guiding treatment decisions.\cite{20hurst2024gpt} By enhancing diagnostic precision and efficiency, EnTao-GPM will be a critical tool in the ongoing evolution of personalized medicine, facilitating better predictions of disease risks and helping optimize treatment strategies based on individual genetic profiles.

% --------------------------------------------------
\section{Method}

\subsection{Model}

Effective pathogenic mutation prediction demands features that capture nucleotide regularities, species context, and clinically verified mutation effects. Inspired by the TrinityDNA framework, we first learn generic sequence representations, then specialise them through post‑training on human and the other 26 laboratory mammalian genomes to obtain \textbf{TrinityDNA‑LabFauna}. Based on this, our new model \textbf{EnTao‑GPM} is fine‑tuned with clinically annotated data, yielding probabilistic pathogenicity scores for individual mutations.

\subsubsection{TrinityDNA}

TrinityDNA serves as the core foundational model, meticulously designed for DNA sequence modeling with bioinspired components that lay the groundwork for subsequent specialized training phases. The architecture comprises \textbf{three principal components} to enable comprehensive DNA sequence processing. The model features parameter sharing and bidirectional modeling capabilities, which are essential for extracting meaningful biological information from the complex DNA sequence space.

\paragraph{Groove Fusion Module (GFM).}  
    DNA strands exhibit prominent major and minor grooves. To encode these multi‑scale structural cues, GFM deploys parallel convolutional filters of sizes $3$, $5$, and $7$, with Gaussian Error Linear Unit (GELU) activations. Formally,
    \begin{equation}
    \operatorname{GrooveFusion}(S)=
    \sum_{k \in \{3,5,7\}}
    \mathrm{GELU}\!\bigl(\operatorname{Conv}_{k}(S)\bigr)
    \label{eq:groovefusion}
    \end{equation}
    where $S$ is the input sequence and $\operatorname{Conv}_{k}$ denotes convolution with kernel size $k$. This design aggregates local features across multiple spatial resolutions.

\paragraph{Sliding Multi‑Window Attention (SMWA).}  
    Conventional self‑attention can suffer from locality bias and oversmoothing on long sequences. SMWA alleviates these issues by assigning each attention head a distinct sliding window size, allowing the model to simultaneously attend to short-range regulatory motifs and long‑range chromosomal interactions.

\paragraph{Gated Reverse Complement (GRC).} 
    Using the symmetry of DNA reverse complement, GRC processes the forward strand and its reverse complement in parallel through a shared transformer encoder. The two representations are subsequently fused with a linear gating mechanism, enriching the feature space of the model while maintaining the efficiency of the parameters.

\subsubsection{TrinityDNA-LabFauna}

Building on TrinityDNA, we derive TrinityDNA-LabFauna (Lab means laboratory focused while Fauna means broad spectrum of experimental mammals) by post‑training the base model on a large‑scale corpus of 27 mammalian genomes(including human and the other 26 laboratory mammals). The training set comprises high‑quality assemblies downloaded from the National Center for Biotechnology Information (NCBI) and spans key model organisms  (Section \ref{subsec:Data}), which dominate contemporary genetics and disease model research. Employing a self‑supervised learning strategy, the model autonomously captures mammal‑specific regularities, such as conserved sequence blocks and the distinction between coding and noncoding regions, without the aid of manual labels. Hyperparameters are re-tuned to emphasize these characteristics of mammals, and the context window is expanded to encompass long-range regulatory architecture and multigene loci (Section \ref{subsec:Training}). This stage equips TrinityDNA-LabFauna with species-aware representations that underpin subsequent tasks, including pathogenic risk prediction.

% \subsubsection{EnTao-GPM}

% The EnTao‑GPM model is obtained by task-specific fine-tuning of TrinityDNA-LabFauna for pathogenic mutation prediction. Fine‑tuning employs two curated resources: the latest ClinVar release available at the time of training, which supplies graded pathogenicity annotations with review confidence, and HGMD, which catalogs only confirmed pathogenic mutations (dataset details in Section \ref{subsec:Data}). Starting from TrinityDNA-LabFauna pre-trained representations, we attach a lightweight classification head and optimize it using an appropriate pathogenicity-weighted loss (hyperparameters in Section \ref{subsec:Training}) to map sequence variations to disease probabilities. After training, EnTao‑GPM outputs the probability that a given mutation is disease‑causing, providing a practical decision aid for geneticists and clinicians.

\subsubsection{EnTao-GPM}

The EnTao-GPM models, including EnTao-GPMFast and EnTao-GPMPro(Section \ref{subsubsec:stage2}), are obtained by task-specific fine-tuning of TrinityDNA-LabFauna for pathogenic mutation prediction. Fine-tuning employs two curated resources: the latest ClinVar release, which provides graded pathogenicity annotations with review confidence, and HGMD, which catalogs only confirmed pathogenic mutations (dataset details in Section \ref{subsec:Data}). Starting from the pre-trained representations of TrinityDNA-LabFauna, we attach a Feedforward Neural Network (FFN) to map sequence variations to disease probabilities. To improve computational efficiency during fine-tuning, we adopt the Low-Rank Adaptation (LoRA)\cite{hu2022lora} technique, which allows for parameter-efficient training by introducing low-rank decomposition during weight updates. After training, EnTao-GPM outputs the probability that a given mutation is disease-causing, providing a practical decision aid for geneticists and clinicians.

\paragraph{Low-Rank Adaptation (LoRA).}
Fine-tuning all parameters in large pre-trained models can be computationally expensive, particularly as model sizes increase. LoRA addresses this challenge by decomposing the weight changes \(\Delta W\) into two smaller matrices \(\Delta W = BA\), where \(B \in \mathbb{R}^{m \times r}\) and \(A \in \mathbb{R}^{r \times n}\), \(r \ll m, r \ll n\), and \(W_1 = W_0 + \Delta W\). This decomposition reduces the number of trainable parameters from \(m \times n\) to \(r \times (m + n)\), significantly improving training efficiency and reducing memory usage without compromising performance. By incorporating LoRA into the fine-tuning process, we ensure that EnTao-GPM is computationally efficient and scalable, enabling it to handle large datasets effectively.

\subsection{Data}\label{subsec:Data}

EnTao-GPM relies on multiple high-quality genomic datasets for its pre-training and fine-tuning processes. The core datasets used to train and evaluate the model include ClinVar, HGMD, and a large collection of genomic data from a variety of species, all contributing to a comprehensive understanding of genetic mutations and their associated risks.

\subsubsection{DNA Pre‑Training Data}\label{subsec:pretrain_data}

To adapt the TrinityDNA backbone to laboratory mammals, we assembled a corpus of 27 complete genomes together with their corresponding cDNA sequences, selecting the experimental species most widely used in genetics and disease research (shown in Table \ref{tab:pretrain_genomes}). Genomic assemblies were obtained from NCBI, and the matching cDNA records were downloaded from the Ensemble repository. The collection spans rodents, primates, livestock, carnivores, and several specialised disease models, amounting to roughly 80B of nucleotide data. Exposure to this diverse sequence set allows TrinityDNA-LabFauna to acquire species specific representations that are essential for downstream pathogenic risk prediction.

\begin{table*}[htbp]
\centering
\small % Reduce the font size for better fitting
\renewcommand{\arraystretch}{1.2} % Adjust row height for compactness
\begin{adjustbox}{max width=\textwidth} % Makes the table fit the page width
\begin{tabular}{|p{4cm}|p{4.5cm}|c|c|c|}
\hline
 \textbf{Name} & \textbf{Ensembl Name} & \textbf{Taxonomy ID} & \textbf{GenBank Assembly ID (GCA)} & \textbf{RefSeq Assembly ID (GCF)} \\
\hline
Homo sapiens & homo\_sapiens & 9606 & GCA\_000001405.29 & GCF\_000001405.40 \\
\hline
 Mus musculus & mus\_musculus & 10090 & GCA\_000001635.9 & GCF\_000001635.27 \\
\hline
Rattus norvegicus & rattus\_norvegicus & 10116 & GCA\_036323735.1 & GCF\_036323735.1 \\
\hline
Macaca mulatta & macaca\_mulatta & 9544 & GCA\_003339765.3 & GCF\_003339765.1 \\
\hline
 Sus scrofa domesticus & sus\_scrofa\_domesticus & 9825 & GCA\_017957985.1 & - \\
\hline
Canis lupus familiaris & canis\_lupus\_familiaris & 9615 & GCA\_011100685.1 & GCF\_011100685.1 \\
\hline
 Callithrix jacchus & callithrix\_jacchus & 9483 & GCA\_011100555.2 & GCF\_011100555.1 \\
\hline
 Oryctolagus cuniculus & oryctolagus\_cuniculus & 9986 & GCA\_964237555.1 & GCF\_964237555.1 \\
\hline
 Mustela putorius furo & mustela\_putorius\_furo & 9669 & GCA\_011764305.2 & GCF\_011764305.1 \\
\hline
Capra hircus & capra\_hircus & 9925 & GCA\_001704415.2 & GCF\_001704415.2 \\
\hline
Bos taurus & bos\_taurus & 9913 & GCA\_002263795.4 & GCF\_002263795.3 \\
\hline
 Equus caballus & equus\_caballus & 9796 & GCA\_041296265.1 & GCF\_041296265.1 \\
\hline
Ovis aries & ovis\_aries & 9940 & GCA\_016772045.2 & GCF\_016772045.2 \\
\hline
Felis catus & felis\_catus & 9685 & GCA\_018350175.1 & GCF\_018350175.1 \\
\hline
Cavia porcellus & cavia\_porcellus & 10141 & GCA\_034190915.1 & GCF\_034190915.1 \\
\hline
Cricetus cricetus & cricetus\_cricetus & 10034 & GCA\_964304595.1 & - \\
\hline
Mesocricetus auratus & mesocricetus\_auratus & 10036 & GCA\_017639785.1 & GCF\_017639785.1 \\
\hline
Nothocricetulus migratorius & nothocricetulus\_migratorius & 3122392 & - & - \\
\hline
Heterocephalus glaber & heterocephalus\_glaber & 10181 & GCA\_000247695.1 & GCF\_000247695.1 \\
\hline
Neovison vison & neovison\_vison & 452646 & GCA\_020171115.1 & GCF\_020171115.1 \\
\hline
Tupaia belangeri & tupaia\_belangeri & 37347 & GCA\_000181375.1 & - \\
\hline
Chinchilla & chinchilla & 10151 & GCA\_000276665.1 & GCF\_000276665.1 \\
\hline
Pallid bat & pallid\_bat & 9440 & GCA\_027563665.1 & - \\
\hline
Hoary bat & hoary\_bat & 257879 & GCA\_011751065.4 & - \\
\hline
Abo bat & abo\_bat & 258909 & - & - \\
\hline
Red bat & red\_bat & 258930 & GCA\_004026805.1 & - \\
\hline
Sind bat & sind\_bat & 568178 & - & - \\
\hline
\end{tabular}
\end{adjustbox}
\caption{Species and Genomic Data for DNA Pre-Training}
\label{tab:pretrain_genomes}
\end{table*}

% The DNA pre-training phase of EnTao-GPM uses high-quality genomic data sourced from the National Center for Biotechnology Information (NCBI). This dataset covers key mammalian species frequently used in genetic research, including humans, pigs, and mice. These species represent common models in genetic studies and disease modeling. The dataset, which includes approximately 27 genomic sequences and corresponding cDNA sequences, provides a broad range of sequence information that is crucial for learning the underlying structures and variations within DNA. By leveraging this data, EnTao-GPM can develop a deep understanding of general genomic features such as conserved regions, coding sequences, and non-coding regions, which play vital roles in predicting disease risk based on genetic mutations.

\subsubsection{ClinVar Data}

To obtain the EnTao‑GPM model through fine‑tuning of TrinityDNA‑LabFauna, we used the latest ClinVar\cite{7Landrum2025clinvar} release available at the time of training (April 9, 2025), which is a comprehensive repository of human genetic variations annotated with clinical significance. The raw download contains extensive metadata, including graded pathogenicity labels and review confidence scores. After a systematic curation and cleaning process, we retained only high-confidence variants that are unambiguously classified as either “pathogenic/likely pathogenic” or “benign/likely benign.” The resulting dataset comprises approximately 300,000 single-nucleotide variants (SNVs) and 35,000 non-SNVs, \textbf{totaling 338,916 variants}. By focusing on variants with well-documented clinical outcomes, we ensure that EnTao-GPM is trained and evaluated on reliable, clinically relevant evidence, thereby enhancing its robustness in pathogenicity assessment.

\begin{figure}[htbp]
  \centering
  \includegraphics[width=1.0\textwidth]{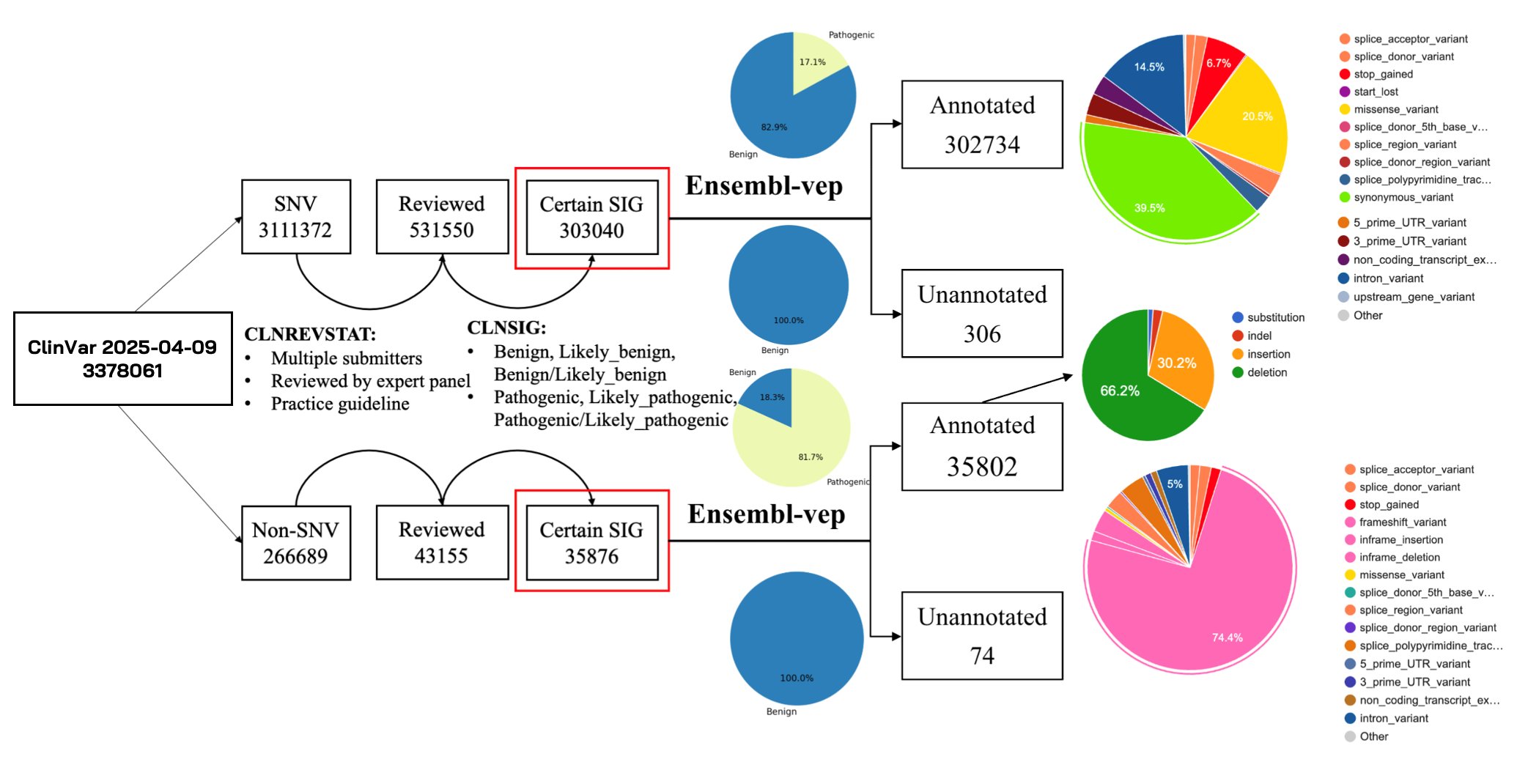}
  \caption{Overview of the curated ClinVar dataset used for fine‑tuning.}
  \label{fig:clinvar}
\end{figure}

% For the fine-tuning of EnTao-GPM, a primary data source is the ClinVar dataset, a comprehensive collection of genetic mutations and their associated pathogenicity annotations. The version used for training EnTao-GPM was downloaded in April 2025 and includes over 300,000 single nucleotide mutations (SNVs) and around 35,000 non-SNVs, such as insertions and deletions. ClinVar provides detailed annotations about each genetic mutation, including disease associations and pathogenicity labels, which are critical for training the model to assess the likelihood that a particular mutation will cause disease.

% The data from ClinVar is carefully curated, with a focus on high-confidence Mutation. The dataset is divided into various categories, including "coding" and "non-coding" mutations, as well as "uncertain" and "benign" classifications. This allows EnTao-GPM to focus on the most clinically relevant and well-supported data, ensuring that the model learns from reliable sources. Importantly, the dataset is rigorously cleaned to exclude ambiguous or low-confidence annotations, ensuring that only mutations with well-documented pathogenicity are used during training.

\subsubsection{HGMD Data}

To complement the ClinVar corpus, we integrated the latest release of the HGMD (professional 2025.2)\cite{11stenson2020human}. Since HGMD exclusively documents disease-causing variants, we restricted our analysis to the "DM" (disease-causing mutation) category, which demonstrates a high degree of concordance (97.2\%) with ClinVar's pathogenic and likely pathogenic labels (shown in Table \ref{tab:consistency_comparison}, we provide an additional Table \ref{tab:mutation_categories} from \cite{11stenson2020human} that offers a detailed description of each HGMD category listed to explain the entries in the first column of Table \ref{tab:consistency_comparison}). In contrast, other HGMD categories show significantly lower levels of agreement with ClinVar's annotations. After carefully removing the overlapping variants with ClinVar, the remaining dataset from the HGMD subset contained \textbf{266,044 variants}. This subset was then merged with the curated ClinVar data to form the final training set. The integrated dataset comprises approximately 290,000 single-nucleotide variants (SNVs) and roughly 28,000 non-SNV mutations. This expanded mutational spectrum not only enhances the diversity of the training data but also ensures the reliability of the pathogenicity labels used for fine-tuning EnTao‑GPM.

\begin{figure}[htbp]
  \centering
  \includegraphics[width=0.9\textwidth]{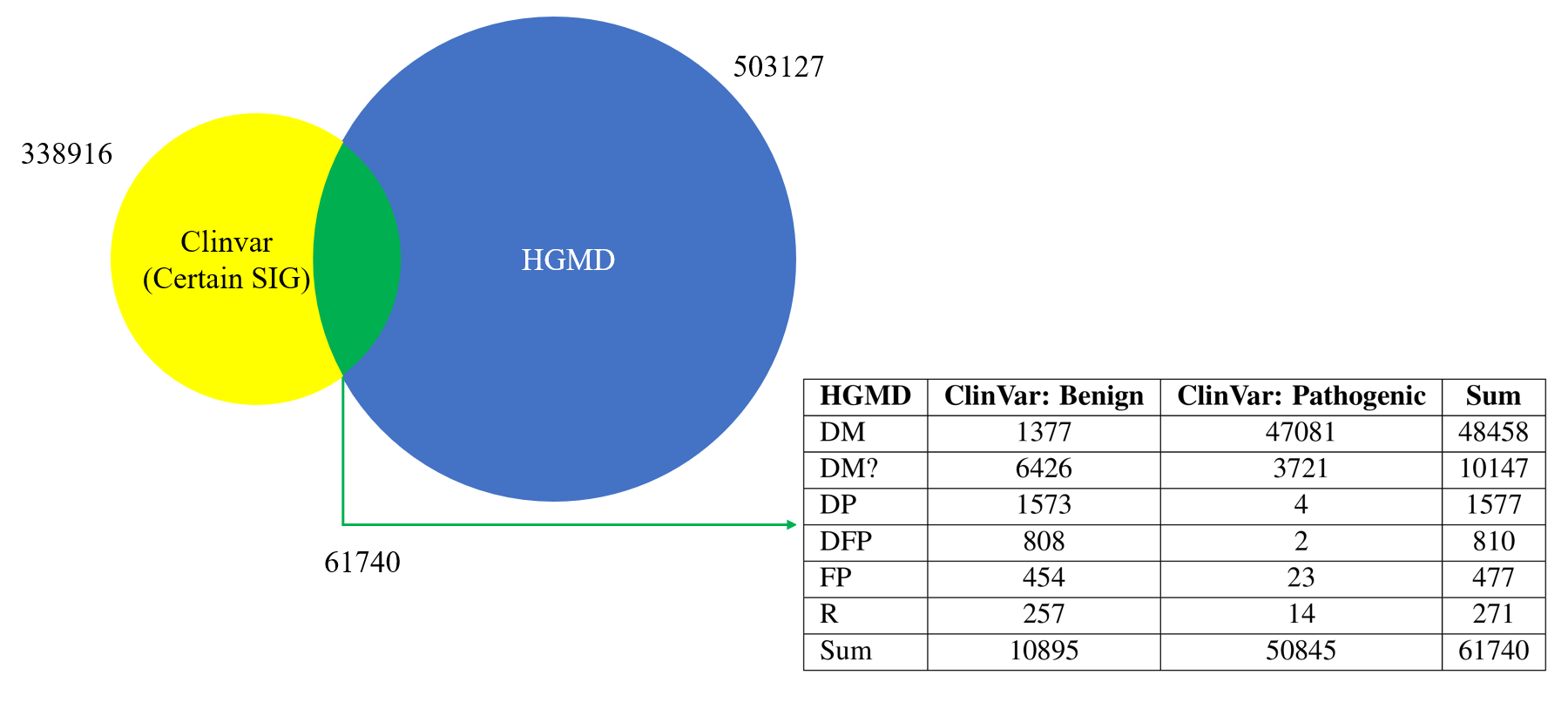}
  \caption{Overview of the HGMD dataset used for fine‑tuning.}
  \label{fig:hgmd}
\end{figure}

% \begin{table*}[htbp]
% \centering

% \begin{tabular}{|l|c|c|c|} % Added | for full borders
% \hline
% \textbf{HGMD} & \textbf{ClinVar: Benign} & \textbf{ClinVar: Pathogenic}& \textbf{Sum} \\
% \hline
% DM & 1377 & 47081 & 48458 \\
% \hline
% DM? & 6426 & 3721 & 10147\\
% \hline
% DP & 1573 & 4 & 1577\\
% \hline
% DFP & 808 & 2 & 810 \\
% \hline
% FP & 454 & 23 & 477 \\
% \hline
% R & 257 & 14 & 271 \\
% \hline
% Sum & 10895 & 50845 & 61740 \\
% \hline
% \end{tabular}
% \caption{Comparison of HGMD and ClinVar Data Consistency}
% \label{tab:consistency_comparison}
% \end{table*}

\begin{table*}[htbp]
\centering
\begin{tabular}{l p{10cm}}
\toprule
\textbf{HGMD variant class} & \textbf{Definition} \\
\midrule
DM & Disease-causing mutation: Literature indicates causal (or likely causal) link with disease. \\
\midrule
DM? & Likely disease-causing, but with additional uncertainty: As for DM, but the authors, curators, or other literature evidence indicate that further caution is warranted. \\
\midrule
DP & Disease-associated polymorphism: Significant statistical association with a clinical phenotype. Likely to be functionally relevant. \\
\midrule
DFP & DP with supporting functional evidence: As for DP, but definitive functional evidence exists (e.g., via an in vitro luciferase assay). \\
\midrule
FP & Functional polymorphism with no reported disease association: Functional effect has been demonstrated, but no disease association has been reported as yet. \\
\midrule
R & Retired from HGMD: Record has been retired and is no longer considered to be phenotypically relevant. \\
\bottomrule
\end{tabular}
\caption{Description of Mutation Categories\cite{11stenson2020human}}
\label{tab:mutation_categories}
\end{table*}

% To supplement the ClinVar dataset, EnTao-GPM also incorporates the latest version of the HGMD (Human Gene Mutation Database) dataset. This dataset provides additional high-confidence data for pathogenic mutations, particularly from the HGMD subset, which specifically includes disease-causing mutations. While there is limited overlap between ClinVar and HGMD, the inclusion of HGMD helps to diversify the model's training data by introducing mutations that are not present in ClinVar but have strong disease associations.

% Data from the HGMD dataset aligns well with ClinVar annotations, especially in cases where mutations are classified as disease-causing. However, variations in labeling across the datasets necessitate careful integration to maintain data consistency. After merging the ClinVar and HGMD datasets, EnTao-GPM was trained on a final curated dataset that includes around 290,000 SNVs and 28,000 non-SNV mutations, covering a wide range of genetic alterations.

\subsubsection{Fine-Tuning Data Preprocessing}

In this study, we prepared two versions of the dataset: \textbf{Dataset 1} and \textbf{Dataset 2}. \textbf{Dataset 1} consists solely of SNVs, while \textbf{Dataset 2} includes both SNVs and non-SNVs.

For \textbf{Dataset 1}, we used a total of 303,040 SNV entries from ClinVar. From this dataset, we selected the highest-confidence variants from the CLNSIG annotation, which were classified as either "Benign" or "Pathogenic" (5406 variants) for testing, referred to as ClinVar SNV Test. The remaining SNV data were used for training, labeled as ClinVar SNV Train.

For \textbf{Dataset 2}, we used the ClinVar SNV Train data from Dataset 1 for training and the ClinVar SNV Test for testing. In addition, to complement the non-SNV data, we selected 10\% of the ClinVar Non-SNV dataset (35876 variants) for testing, referred to as ClinVar Non SNV Test, while the remaining data was used for training (ClinVar Non SNV Train). Furthermore, we incorporated the HGMD dataset to include additional disease-causing variants. From this dataset, 2000 variants were randomly selected for testing (1,543 SNVs and 457 non-SNVs), referred to as HGMD Test. The remaining data from HGMD was used for training.

The final dataset compositions are shown in Table \ref{tab:dataset_composition}. These datasets were used to fine-tune the EnTao-GPM model for genetic mutation pathogenicity prediction.

\begin{table*}[htbp]
\centering
\small % Reduces the font size
\renewcommand{\arraystretch}{1.2} % Adjusts the row height to make it more compact
\begin{tabular}{|l|p{5cm}|p{5cm}|l|}
\hline
\textbf{Dataset} & \textbf{Training Data} & \textbf{Test Data} & \textbf{Mutation Type} \\
\hline
Dataset 1 & ClinVar SNV Train (297634) & ClinVar SNV Test (5406) & SNV \\
\hline
Dataset 2 & ClinVar SNV Train (297634) & ClinVar SNV Test (5406) & SNV \\
& ClinVar Non-SNV Train (32295) & ClinVar Non-SNV Test (3581) & Non-SNV \\
& HGMD Train (264044) & HGMD Test (2000) & SNV and Non-SNV \\
\hline
\end{tabular}
\caption{Dataset Composition for Fine-Tuning}
\label{tab:dataset_composition}
\end{table*}

% \subsubsection{Model Training and Fine-Tuning}

% The model’s training process is divided into two key stages: the DNA pre-training and the fine-tuning phases. During DNA pre-training, EnTao-GPM learns to identify and represent the core features of DNA sequences through unsupervised learning techniques applied to the diverse genomic data. This phase allows the model to build a foundational understanding of genomic structures that is independent of specific disease annotations.

% In the fine-tuning phase, EnTao-GPM uses the curated ClinVar and HGMD datasets to adapt the general genomic representations learned during pre-training to specific clinical tasks. This supervised learning phase enables the model to learn the relationships between genetic mutations and their pathogenicity, producing highly accurate predictions of disease risk based on genetic data. The fine-tuning process involves optimizing model parameters through a series of epochs, with performance evaluated on validation and test sets.

% --------------------------------------------------
\section{Experiment}

\subsection{Training}\label{subsec:Training}
% To be added upon release of training schedule, hyperparameters, and strategies.

\subsubsection{Stage 1: DNA Pre-Training}
The pre-training of TrinityDNA-LabFauna builds upon the foundational architecture and training paradigm of TrinityDNA \cite{yang2025trinitydna}, with a focus on adapting the model to laboratory mammal genomes. This stage employs an evolutionary training strategy, inspired by TrinityDNA's two-phase approach, but specializes in mammalian genomic features.

Adhering to TrinityDNA's self-supervised learning framework, we employed masked language modeling (MLM) as the primary pre-training objective \cite{devlin2019bert}. This strategy helps the model learn contextual dependencies and generalize across unseen sequences.

Training was performed using the same framework as TrinityDNA, built on Megatron and DeepSpeed with FlashAttention 2 for efficient computation \cite{yang2025trinitydna}. We employed 4D parallelism (Pipeline, Model, Data, and DeepSpeed-Ulysses) to handle large sequence lengths, with hyperparameters fine-tuned for mammalian data. Specifically, the context window was set to 100k base pairs to accommodate long regulatory regions in eukaryotic genomes. The model was trained on A100 GPUs with BF16 precision for acceleration and FP32 gradient accumulation for numerical stability.

This pre-training phase enables TrinityDNA-LabFauna to develop a deep understanding of mammalian genomic characteristics, including conserved regions, coding/non-coding distinctions, and species-specific regulatory elements, providing a solid foundation for fine-tuning on pathogenic risk prediction tasks.

\subsubsection{Stage 2: Mutation Prediction Fine-Tuning}\label{subsubsec:stage2}

\begin{figure}[htbp]
\centering
\includegraphics[width=0.8\textwidth]{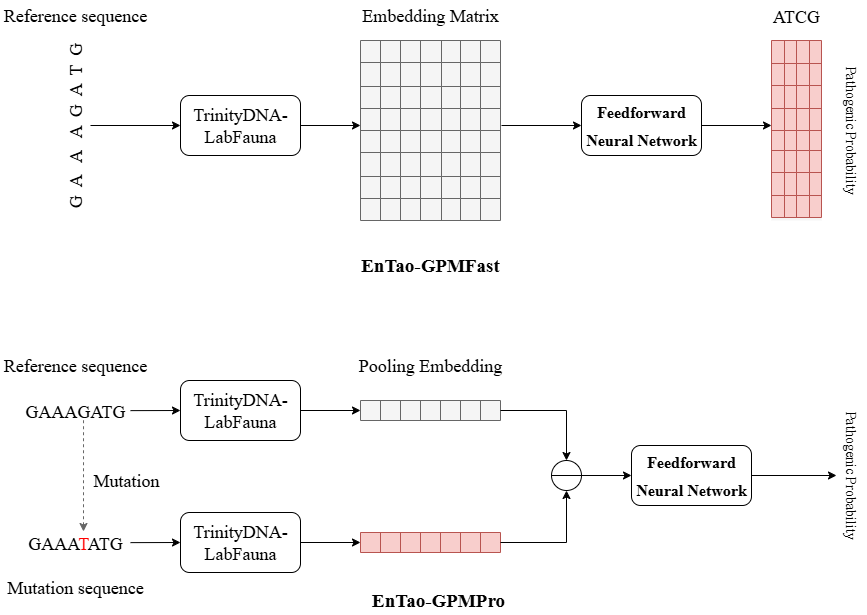}
\caption{EnTao-GPM model architecture for mutation prediction.}
\label{fig:model_architectures}
\end{figure}

In this stage, we fine-tuned the TrinityDNA-LabFauna model using LoRA\cite{hu2022lora} to adapt it for mutation prediction tasks, utilizing two distinct model architectures: EnTao-GPMFast and EnTao-GPMPro. Both models were trained on different datasets, each tailored for specific mutation types.

\textbf{EnTao-GPMFast} is designed to predict the pathogenicity of SNV mutations. It operates solely on the reference genome sequence, enabling rapid mutation pathogenicity predictions through Disease Mutation Scanning (DMS). By directly inputting the reference genome sequence, EnTao-GPMFast can efficiently predict the pathogenicity of a large number of point mutations. For broader applicability, we performed a whole-genome mutation pathogenicity scan on the human reference genome (GRCh38 version) using EnTao-GPMFast, and the results were made publicly available for rapid mutation prediction across multiple scenarios.

\textbf{EnTao-GPMPro}, in contrast, requires both the reference genome sequence and the corresponding mutation sequence. It is designed to predict the pathogenicity of specific mutations and can handle both SNV and Non-SNV mutations, making it suitable for more complex mutation scenarios.

For training, we used the training set from DataSet1 to train EnTao-GPMFast, while EnTao-GPMPro was trained on the training set from DataSet2. The specific details of the training and test sets used are provided in Table ~\ref{tab:dataset_composition}.

\subsection{Results}

\subsubsection{DNA Pre-Training}
The pre-training of TrinityDNA-LabFauna was conducted using a model architecture derived from TrinityDNA, trained on the genomic data of 27 species. This model was trained following the same method as TrinityDNA, utilizing genomic data from species widely used in genetics and disease research. The performance of TrinityDNA-LabFauna was evaluated on the ClinVar-SNV test dataset using a zero-shot approach, where the model was directly applied without fine-tuning.

The results, as shown in Table~\ref{tab:comparison_results}, demonstrate that TrinityDNA-LabFauna outperforms the EVO2-1B model with the same parameter size, and achieves comparable performance to the EVO2-40B model, which has the largest number of parameters. Specifically, TrinityDNA-LabFauna achieves an \textit{AUC} of 0.943 and an \textit{AUPRC} of 0.961, outperforming the EVO2-1B model in both metrics (\textit{AUC}: 0.950, \textit{AUPRC}: 0.955). Moreover, TrinityDNA-LabFauna's performance is closely aligned with the highest-performing model, EVO2-40B, indicating that it has successfully captured complex genomic dependencies during pre-training.

\begin{table*}[htbp]
\centering
\small % Reduce the font size further for better fitting
\renewcommand{\arraystretch}{1.3} % Adjust row height to be more compact
\begin{adjustbox}{max width=\textwidth} % Makes the table fit the page width
\begin{tabular}{|l|l|r|r|r|r|r|r|}
\hline
\textbf{Dataset} & \textbf{Metric} & \textbf{FudanDNA-1B} & \textbf{AIDO.DNA-7b} & \textbf{MambaDNA} & \textbf{EVO2-40b} & \textbf{EVO2-7b} & \textbf{EVO2-1b-base} \\
\hline
ClinVar & \textit{AUC} & \underline{0.943} & 0.547 & 0.672 & \textbf{0.950} & 0.945 & 0.927 \\
ClinVar & \textit{AUPRC} & \textbf{0.961} & 0.621 & 0.735 & \underline{0.955} & 0.957 & 0.946 \\
\hline
\end{tabular}
\end{adjustbox}
\caption{Performance Comparison of TrinityDNA-LabFauna with Other Models on ClinVar Dataset}
\label{tab:comparison_results}
\end{table*}

\subsubsection{Mutation Prediction Fine-Tuning}

For the EnTao-GPMFast model, which only supports SNV mutation pathogenicity prediction, we evaluated its performance using the DataSet 1 test set. As shown in the results, EnTao-GPMFast performs better in the SNV scenario with an \textit{AUC} of 0.963. However, its architecture does not support Non-SNV mutation prediction. In this regard, we utilized the original pre-trained model, TrinityDNA-LabFauna. In this approach, we applied the pre-trained model with masked mutation positions and inferred the perplexity for both the mutated and wild-type sequences. The difference between the perplexity of the wild-type and mutated sequences was used as the zero-shot prediction result. This method effectively simulates a prediction without fine-tuning, showcasing the baseline model performance. TrinityDNA-LabFauna, as expected, shows a significant drop in performance in Non-SNV scenarios, where it is unable to handle the complex mutation types effectively.

In contrast, the EnTao-GPMPro model, specifically designed and fine-tuned for Non-SNV mutation prediction, retains excellent performance even on Non-SNV data, with an \textit{AUC} of 0.933. The fine-tuning was conducted using DataSet 2, which includes both SNV and Non-SNV mutations, allowing EnTao-GPMPro to learn complex patterns and provide accurate predictions for a broader set of mutation types.

From a practical perspective, EnTao-GPMFast is designed to be simple and fast for inference, requiring only the reference genome as input, making it ideal for large-scale batch evaluations of the pathogenicity of SNV. Its ability to rapidly process SNV predictions is a major advantage in large-scale genomic screening tasks. On the other hand, EnTao-GPMPro is designed for more complex mutation scenarios, supporting both SNV and Non-SNV mutations and enabling predictions of combined effects in multipoint mutations. This capability is critical for applications where a broader spectrum of mutation types must be considered, such as in genetic diseases involving single nucleotide variations and structural variants.

The fine-tuning results are summarized in Table~\ref{tab:finetuning_results}. As shown, the EnTao-GPMPro model performs well across both the SNV and Non-SNV test sets, providing a robust solution for diverse mutation types. In contrast, EnTao-GPMFast excels only in the SNV context, demonstrating the specialized nature of its architecture.

\begin{table*}[htbp]
\centering
\small % Reduce font size further for better fitting
\renewcommand{\arraystretch}{1.2} % Adjust row height for compactness
\begin{adjustbox}{max width=\textwidth} % Makes the table fit the page width
\begin{tabular}{|l|l|c|l|c|r|r|}
\hline
\multicolumn{7}{|c|}{\textbf{(a) Performance on DataSet 1 Test}} \\
\hline
\textbf{Model} & \textbf{Training Dataset} & \textbf{Training Data Size} & \textbf{Test Dataset} & \textbf{Test Data Size} & \textbf{AUC} & \textbf{AUPRC} \\
\hline
Zero-shot & No & - & DataSet 1 Test & 5406 & 0.943 & 0.961 \\
EnTao-GPMFast & DataSet 1 Train & 297634 & DataSet 1 Test & 5406 & 0.963 & 0.974 \\
EnTao-GPMPro & DataSet 2 Train & 593973 & DataSet 1 Test & 5406 & 0.960 & 0.970 \\
\hline
\multicolumn{7}{|c|}{\textbf{(b) Performance on DataSet 2 Test}} \\
\hline
Zero-shot & No & - & DataSet 2 Test & 10987 & 0.704 & 0.885 \\
EnTao-GPMFast & DataSet 1 Train & 297634 & DataSet 2 Test & 10987 & - & - \\
EnTao-GPMPro & DataSet 2 Train & 593973 & DataSet 2 Test & 10987 & 0.933 & 0.966 \\
\hline
\end{tabular}
\end{adjustbox}
\caption{Fine-Tuning Results on DataSet 1 and DataSet 2}
\label{tab:finetuning_results}
\end{table*}

% The EnTao-GPM model has been comprehensively evaluated on the ClinVar benchmark dataset using zero-shot and fine-tuned settings. In the zero-shot scenario, the model demonstrated robust generalization capability, achieving an AUC of 0.960 and AUPRC of 0.970 for SNV classification, outperforming the existing xTrimoDNA-1B model and matching the performance of models with significantly larger parameter scales such as EVO2-40B. The model was further fine-tuned on subsets of SNV-only, non-SNV-only, and mixed SNV/non-SNV data. For each configuration, performance was assessed via area under the ROC and PR curves, as well as epoch-wise accuracy and loss convergence on separate validation and test sets. Results show that including both SNV and non-SNV data in training yields more balanced and generalizable predictions across mutation types.

% Additional evaluations were conducted on the HGMD test set to assess cross-database generalization. Despite limited overlap between HGMD and ClinVar, models trained on the combined dataset maintained strong performance on ClinVar and exhibited improved prediction accuracy on HGMD-exclusive Mutation. This finding suggests that integrating diverse data sources with consistent pathogenic annotations can enhance model robustness in real-world applications.

% --------------------------------------------------
\section{Conclusion}

In this study, we introduced \textbf{EnTao-GPM}, a cutting-edge model for the prediction of germline pathogenic mutations. The model integrates advanced DNA sequence pre-training with fine-tuning on clinically validated mutation datasets, making it a robust tool for genetic mutation interpretation. EnTao-GPM leverages the power of the TrinityDNA framework, which has been extended and fine-tuned on genomic data from 27 species to capture species-specific genomic patterns critical for pathogenicity prediction.

The EnTao-GPM model has demonstrated superior performance in distinguishing pathogenic from benign mutations in various genomic contexts. Specifically, it outperforms existing models, such as EVO2-1B, in predicting SNVs and exhibits comparable performance to the largest models available, including EVO2-40B. Moreover, its fine-tuning on both SNV and non-SNV data allows it to provide accurate pathogenicity predictions in more complex mutation scenarios, such as those involving insertions and deletions.

The results from the fine-tuning phase highlight the practical applications of the model. EnTao-GPMFast, which is optimized for rapid SNV predictions using only the reference genome, provides fast, large-scale predictions for SNVs. On the other hand, EnTao-GPMPro supports both SNV and non-SNV mutations, making it suitable for more intricate mutation scenarios that are typically encountered in complex genetic diseases. The fine-tuning results demonstrate that EnTao-GPMPro excels in handling a wide range of mutation types, providing robust solutions for clinical genomic diagnostics.

% Section reserved for discussion of model generalization limits, rare mutation interpretation challenges, and future expansion plans.

% --------------------------------------------------
\section{Future Work}

The promising results achieved by EnTao-GPM in pathogenic mutation prediction provide a solid foundation for expanding its capabilities in several key directions.

\paragraph{Integration of Multi-Omics Data and Knowledge Bases.}
Future work will focus on enhancing EnTao-GPM by integrating multi-omics data and authoritative knowledge bases like OMIM. This integration will enable more comprehensive predictions by linking genomic variants to phenotype information, thus improving the model’s ability to interpret rare and complex mutations. Additionally, incorporating other omics layers, such as transcriptomics and epigenomics, will allow the model to better capture the regulatory context of mutations, especially for non-coding regions and complex diseases like cancer.

\paragraph{Fusion of Sequence Models and Language Models for Contextual Reasoning.}
To bridge the gap between genomic data and clinical applications, we aim to combine EnTao-GPM with large language models (LLMs) trained on biomedical literature. This fusion will allow the model to provide natural language explanations for pathogenicity predictions and contextualize mutation impacts in the context of clinical data, such as patient history and disease comorbidities. The ability to interpret complex cases, such as polygenic disorders, will make EnTao-GPM more accessible for clinical decision support.

\paragraph{Expansion to Somatic Mutations and Cancer Susceptibility.}
The subsequent study will develop a dedicated somatic-variation module that complements the present germline framework. This line of research will leverage cancer‑focused resources, such as COSMIC and TCGA—and devise representation strategies for structural variants (SVs) and copy‑number variations (CNVs). By jointly modelling germline and somatic alterations, the resulting system is expected to deliver finer cancer‑risk stratification and more reliable discrimination between driver and passenger mutations in tumour genomes.

% \section{Application Scenarios}

% \subsection{Application}

% The EnTao-GPM model is expected to have wide-ranging applications in clinical genomics, personalized medicine, and therapeutic development. In clinical settings, EnTao-GPM can function as a decision support tool for interpreting genetic sequencing results, enabling healthcare providers to assess mutation pathogenicity with higher accuracy and consistency. It provides actionable insights for patient-specific disease risk evaluation and can assist in early-stage diagnosis of monogenic disorders. From a translational research perspective, the model facilitates the identification of potentially disease-driving mutations, supporting drug target prioritization and biomarker discovery. Moreover, its capability to process non-SNV mutations makes it suitable for comprehensive genome interpretation workflows. The model’s integration with Python-based APIs, VCF pipelines, and GRCh38 genome standards ensures compatibility with common clinical bioinformatics platforms.

\section{Inference Interface}

To support deployment across diverse clinical and research environments, EnTao-GPM offers multiple inference modes. The model provides a Python-based local library and a web-accessible server API. Inputs can be supplied in standard formats such as VCF or CSV, and the model outputs mutation-level pathogenicity probabilities. It supports both SNV inputs mapped to GRCh38 and non-SNV sequences beyond standard genome coordinates. Inference can be conducted on local GPU servers with a minimum of 40GB of memory or cloud-based endpoints, with performance optimized for rapid batch processing and low-latency clinical usage.
% --------------------------------------------------
\section{Responsible Release and Access}

For technical inquiries or research collaboration, please contact Fan Zhong (\url{zhongfan@fudan.edu.cn}). The EnTao-GPM project is a collaborative initiative led by Fudan University and BioMap, aiming to develop vertical AI systems that meet clinical-grade robustness and interpretability standards.

% Section reserved for licensing, distribution strategy, and usage boundaries.

% --------------------------------------------------
\section{Funding}
This work was supported by the National Key Research and Development Program of China (2024YFA1307702), the Shanghai Science and Technology Innovation Action Plan in Computational Biology (24JS2840200), and the Peak Disciplines (Type IV) of Institutions of Higher Learning in Shanghai.

\section{Acknowledgement}
This work has been supported by the Medical Science Data Center in Shanghai Medical College of Fudan University.

\bibliographystyle{unsrt}  % 格式样式
\bibliography{ref.bib}  % 关联bib文件

% \section*{Appendix}
\end{document}